\documentclass[journal,12pt,onecolumn,draftclsnofoot,]{IEEEtran}
\usepackage{graphicx}
\usepackage{epstopdf}
\usepackage{amsmath,amssymb,amsthm,array}
\usepackage{epsf}
\usepackage{color}
\usepackage{subfigure}
\usepackage[ruled,vlined, linesnumbered]{algorithm2e}
\usepackage{cite}
\usepackage{enumitem}
\usepackage{amsfonts}
\usepackage{comment}
\usepackage{graphicx}
\usepackage{textcomp}
\usepackage{makeidx,epsfig}
\usepackage{diagbox}
\usepackage{multirow}
\usepackage{algpseudocode}
\usepackage{multicol}
\usepackage{verbatim} % multiline comment
\usepackage{cases}
\usepackage{cleveref}
\usepackage{lipsum}

\usepackage{alltt,tabu}
\usepackage{booktabs}
\usepackage{enumerate}
\usepackage[dvipsnames]{xcolor}
\usepackage[utf8]{inputenc}

\def\Phib{{\boldsymbol \Phi}}

\newtheorem{theorem}{\sc Theorem}
\newtheorem{assumption}{Assumption}
\newtheorem{definition}{Definition}

\def\b1{{\bf 1}}

\def\diag{{\rm diag}}

\definecolor{darkgreen}{rgb}{0.0, 0.5, 0.0}

\begin{document}
%\title{Deep Unfolding-Aided Collaborative Learning for Distributed Compressive Phase Retrieval With Multi-View Affine Measurements}
%\title{Collaborative Learning for Distributed Multi-View Compressive Affine Phase Retrieval in Resource-Constrained Wireless Sensor Networks}
%\title{Collaborative Learning for Distributed Compressive Phase Retrieval from Multi-View Measurements}
%\title{Collaborative Learning for Distributed Multi-View Compressive Phase Retrieval With Outliers}
\title{Robust Distributed Multi-View Compressive Phase Retrieval With Outliers}
\title{Robust Distributed Phase Retrieval for Multi-View Compressive Networked Sensing With Outliers }
%\title{Distributed Multi-View Compressive Affine Phase Retrieval in Multi-Sensor IoT Networks} %Resource-Constrained Sensor Networks} 

\author{Ming-Hsun Yang\thanks{This work was supported in part by the National Science and Technology Council (NSTC) of Taiwan under grants NSTC 113-2221-E-008-069, and 112-2221-E-008-057.}
\thanks{M.-H.~Yang is with the Department of Electrical Engineering and the Institute of Computer and Communication Engineering, National Cheng Kung University, Tainan 70101, Taiwan %the Department of Communication Engineering, National Central University, Taoyuan 320, Taiwan
(e-mail: mhyang@cc.ncu.edu.tw).}
%\thanks{Y.-W.~P.~Hong is with the Institute of Communications Engineering, National Tsing Hua University, Hsinchu 30013, Taiwan (e-mail: ywhong@ee.nthu.edu.tw).}
        }
%\name{You-De~Huang,
%        and~Y.-W.~Peter~Hong\thanks{This work was supported in part by the Ministry of Science and Technology, Taiwan, under grant MOST 110-2634-F-007-021.}}
%\address{$^\dag$Institute of Communications Engineering, National Tsing Hua University, Taiwan}

\maketitle

\begin{abstract}
This work examines the multi-view compressive phase retrieval problem in a distributed sensor network, where each sensor device, limited by storage and sensing capabilities, can access only intensity measurements from %a certain (and unknown) part of 
an unknown part of the global sparse vector. The goal is to enable each sensor to recover its observable sparse signal when measurements are corrupted by outliers. To achieve reliable local signal recovery with limited data access, we propose a distributed reconstruction algorithm that enables collaboration among sensor devices without the need to share individual raw data. The proposed scheme employs a two-stage approach that first recovers the amplitude of the global signal (at a central server) and subsequently estimates the observable nonzero signal entries (at each local device). Our analytic results show that perfect global signal amplitude recovery can be achieved under mild conditions on the support size of sparse outliers and the view blockage level. %The exact identification of local observed signal components is also shown to be attainable subject to a mild requirement on the noise support set size by solving a binary optimization problem. 
In addition, the exact reconstruction of locally observed signal components is shown to be attainable in the noise-free case by solving a binary optimization problem, subject to a mild requirement on the structure of the sensing matrix. Computer simulations are provided to illustrate the effectiveness of the proposed scheme. %subject to mild conditions on the support size for sparse outliers and the sparse signal.  %In this problem, sensors aim to recover their respective observation signals in the presence of outlier corruption. To achieve reliable signal recovery despite limited data access and outlier corruption, we propose a distributed reconstruction algorithm that enables collaboration among sensor devices without the need to share individual raw data. The proposed signal reconstruction protocol involves:

%CS is an emerging technique capable of enhancing spectrum utilization for the future internet-of-things (IoT) systems. Wireless sensor networks (WSNs) have proven to be highly effective for inference in a range of modern applications, including environmental monitoring [1], battlefield surveillance [2], and structural health monitoring [3], among others. In the realm of signal processing, two critical techniques that empower WSNs are parameter estimation and signal/event detection. At the same time, emerging applications for WSNs present challenges related to cost-effective sensor deployment and reduced energy consumption. Consequently, the development of low-cost and energy-efficient distributed estimation and detection algorithms for WSNs holds significant importance. This thesis concentrates on the topics of distributed sparse signal retrieval and distributed dynamic event region detection. Our objectives involve designing new signal processing protocols and algorithms, primarily focusing on enhancing energy efficiency, and exploring their attainable performance.
\end{abstract}


% \fi
\IEEEpeerreviewmaketitle

\section{Introduction}
\label{sec:intro}

Wireless sensor networks (WSNs) and the Internet of Things (IoT) consist of a large number of resource-constrained sensors or devices that are designed to sense, collect, and process data from monitored environments.
%Wireless sensor networks (WSNs) and the Internet of Things (IoT) consist of a large number of resource-constrained sensors or devices that are deployed to sense, collect, and process information within monitored environments. 
Over the past decades, these multi-sensor networks have been widely used for inference and detection in various applications \cite{Deligiannis17,WangTian18}. %, including environmental monitoring \cite{Deligiannis17}, battlefield surveillance \cite{YuWang19}, and structural health monitoring \cite{WangTian18}.
To enable massive machine-to-machine communication or real-time industrial monitoring and control within the network,
the algorithm design of efficient data fusion and reconstruction for communication and signal sensing has become an important research topic in recent years.  In particular, the design of resource-constrained WSNs (or IoT systems) often faces the challenge of efficient acquisition, storage, and processing of sensory data \cite{WangTian18}. %To cope with the aforementioned issue, many solutions from a physical-layer signal processing perspective have been devised in the past few decades [2]– [4]. 
As reported in many existing empirical studies \cite{Eldar11}, real-world signals typically lie in low-dimensional subspaces of the ambient domain, thereby admitting a sparse representation under a certain basis. This finding has inspired the development of state-of-the-art compressive sensing (CS) techniques \cite{Eldar11}, as well as other sparsity-promoting schemes, to cope with the aforementioned issue. The sub-Nyquist nature %(i.e., data acquisition at rates far below the Nyquist frequency) 
of CS potentially economizes data gathering and storage; the reduced amount of sampled data can further facilitate efficient signal processing and conserve subsequent data transmission overheads. These benefits make CS particularly attractive for the design of WSNs \cite{Lv23,ChenJY15,tian2023}.  %All these benefits have made CS suitable for the design of WSNs \cite{Lv23,ChenJY15,tian2023}. %and IoT systems [7]–[9].
%within green IoT networks

In the context of compressive WSNs, distributed cooperative sparse signal estimation and reconstruction has received considerable attention in recent years, e.g., \cite{Lv23,ChenJY15}. Most existing works focus on the full-view scenario, which assumes that the entire signal of interest is observable at every node. However, this assumption may be impractical in many sensing applications due to occlusion effects and blind spots caused by the geographic locations of sensors, or limited sensing capabilities resulting from energy constraints. As a result, certain unknown components of the global signal may be missing or unobservable. Such multi-view sensing scenarios arise in numerous engineering applications, including distributed environment sensing in Integrated Sensing and Communications (ISAC) \cite{Tong22} and industrial IoT systems \cite{Cai18}. To address this issue, \cite{tian2023} formulated the problem as a bilinear optimization task based on a factored joint sparsity model, and solved it in a distributed manner using the consensus ADMM algorithm. However, this method requires access to both the magnitude and phase of the measurements, %knowledge of both the magnitude and phase of the measurements, 
which may limit its applicability, %particularly in situations where the
especially when phase information is unavailable or unreliable. %especially when the phase information in the measurements is not available (or unreliable). 
In-depth study on such issues, however, remains limited in the literature.
%In-depth study of issues of this kind, however, remains limited in the existing literature.

In this letter, we examine a multi-view compressive phase retrieval problem in WSNs with outlier corruption. To reduce in-network data acquisition costs, each device employs a sparse sensing matrix to obtain compressed measurements. Exploiting the sparse structure of both the global signal and the sensing matrices, we propose an efficient distributed signal recovery algorithm that enables device collaboration for robust recovery of local signals\footnote{This work primarily focuses on the reconstruction of locally observed signals. An interesting direction for future work is to extend the current framework toward global signal reconstruction in multi-view sensing scenarios.}, without explicitly sharing phaseless multi-view measurements. Specifically, the proposed method follows a two-stage approach: it first performs global signal amplitude reconstruction at a central server, followed by multi-view local signal recovery at each device. Theoretical performance guarantees are established, and numerical simulations are provided to demonstrate the effectiveness of the proposed distributed recovery scheme. %Specifically, the proposed method relies on two-stage approach that consists of the amplitude reconstruction of the global signal (at a central server) followed by multi-view local signal retrieval (at each local device). The analytic performance guarantees are provided. Also, computer simulations are conducted to illustrate the effectiveness of the proposed distributed recovery scheme.    %To reduce the data acquisition cost, each device adopts a sparse sensing strategy by employing a sparse sensing matrix to obtain compressed observations.
%each device in the network adopts a sparse sensing strategy to obtain compressed observations. In the proposed approach, each  %to enable collaboration among devices without explicitly sharing their phaseless measurements.  

\section{Problem Statement and Basic Assumptions}
\label{sec:problem_statment}

%\begin{figure}[t]
%\centering{\includegraphics[width=.8\linewidth]{system2.pdf}}
%\caption{Illustration of the distributed partial-view cooperative learning framework.}
%\label{fig:model}
%\end{figure}

We consider a distributed network with $I$ local edge devices (or IoT nodes) and a central server to coordinate their computations for joint sparse signal reconstruction. The global sparse signal $\mathbf{s}=[s_1\cdots s_N]^T\in\mathbb{R}^N$ is assumed to be supported on the unknown set $\mathcal{K}\subset\{1,\dots,N\}$ with cardinality $|\mathcal{K}|=K$, where $K\ll N$. Each nonzero entry of $\mathbf{s}$ is independently drawn from a continuous probability distribution. %These non-zero entries in $\mathbf{s}$ are independently drawn from continuous probability distributions.
In the multi-view sensing scenario, each local device may not make a full observation of $\mathbf{s}$ due to its limited sensing capability. Hence, the local observable signal at node $i$, $1\leq i\leq I$, is given by   %In many IoT applications, local edge devices may not make a full observation of $\mathbf{s}$ due to their limited sensing capabilities or the presence of blind spots within their respective coverage areas. We therefore consider the scenario in which each local node within the network only observes a certain part of $\mathbf{s}$. The local observable signal at node $i$, $1\leq i\leq I$, can be modeled as  %We assume that the local devices may not make a full observation of the signal $\mathbf{s}$ due to their limited sensing capabilities or the presence of blind spots within their respective coverage areas. We instead consider the scenario in which each local node within the network only observes a certain part of $\mathbf{s}$. Hence, the local observable signal at node $i$, $1\leq i\leq I$, can be modeled as
\begin{align}\label{eq:partial_view_model}
    \mathbf{s}^{(i)}= \mathbf{D}^{(i)}\mathbf{s},
\end{align}
where $\mathbf{D}^{(i)}=\diag(d_1^{(i)},\dots,d_N^{(i)})$ is the (unknown) masking diagonal matrix indicating the observation capability of node $i$. Here, $d_n^{(i)}=1$, if node $i$ is able to make an observation of the $n$th signal component  $s_n$, and $d_n^{(i)}=0$, otherwise. Notably, when node $i$ has full access to $\mathbf{s}$, we have $\mathbf{D}^{(i)}=\mathbf{I}_N$. %This special case, where $\mathbf{D}^{(i)}=\mathbf{I}_N$ holds for all $i$, has been widely studied in the CS literature.   %when node $i$ can obtain a full view of $\mathbf{s}$.
Many existing works in the CS literature have focused on this scenario where $\mathbf{D}^{(i)}=\mathbf{I}_N$ holds for all $i$.  By using random projections for data reduction, the $M_i<N$ compressed observations available at the $i$th local device then obey the following nonlinear measurement model    %%for joint sparse signal reconstruction, as depicted in Fig. ??.
\begin{align}\label{eq:signal_vector_model}
    \mathbf{y}^{(i)}= [y_1^{(i)} \cdots y_{M_i}^{(i)}]^T=\left|\Phib^{(i)}\mathbf{s}^{(i)}\right|^2+\mathbf{w}^{(i)},
\end{align}
where $|\cdot|^2$ denotes the
element-wise absolute-squared value, $\Phib^{(i)}=[(\underline{\boldsymbol{\phi}}_{1}^{(i)})^T \cdots (\underline{\boldsymbol{\phi}}_{M_i}^{(i)})^T]^T\in\mathbb{R}^{M_i\times N}$ is the sensing matrices (with $\underline{\boldsymbol{\phi}}_{m}^{(i)}=[\phi_{m,1}^{(i)} \cdots \phi_{m,N}^{(i)}]\in\mathbb{R}^{1\times N}$ being its $m$th row), %$\mathbf{b}^{(i)}=[b_1^{(i)} \cdots b_{M_i}^{(i)}]^T\in\mathbb{C}^{M_i}$ is the bias vector, 
and $\mathbf{w}^{(i)}=[w_1^{(i)} \cdots w_{M_i}^{(i)}]^T\in\mathbb{R}^{M_i}$ is the sparse noise vector (i.e., outlier) with arbitrary nonzero values. The model considered in \eqref{eq:signal_vector_model} is applicable to atomic MIMO communications, such as channel estimation and signal detection in quantum sensing systems \cite{Cui25,Kim25}.

To facilitate efficient data acquisition and subsequent cooperative learning, each local device exploits a sparse sensing matrix $\Phib^{(i)}$ in \eqref{eq:signal_vector_model} to sketch the signal. Specifically, we adopt the widely used (i.i.d.) \textit{Bernoulli random sensing} framework, where each entry of $\Phib^{(i)}$ independently takes a nonzero value with probability $q$, i.e., $\Pr(\phi_{m,n}^{(i)}\neq 0)=q$ for all $m$, $n$ and $i$. The non-zero entries of $\Phib^{(i)}$ are assumed to be independently drawn from continuous probability distributions. %each entry of $\Phib^{(i)}$ is for each $1\leq m\leq M$, the $m$th measurement captures the We denote $\mathcal{C}_n^{(i)}\subset\{1,\dots,M_i\}$ 
%The support set of column $n$ in $\Phib^{(i)}$ is denoted by $\mathcal{C}_n^{(i)}\subset\{1,\dots,M_i\}$. 
Let $\mathcal{C}_n^{(i)}\subset\{1,\dots,M_i\}$ be the support set of column $n$ in $\Phib^{(i)}$. Then, the measurements $\{y_m^{(i)}\}_{m\in\mathcal{C}_n^{(i)}}$ in \eqref{eq:signal_vector_model} are the only ones capable of capturing the $n$th signal component of $\mathbf{s}^{(i)}$. %Then, it follows from \eqref{eq:signal_vector_model} that only the measurements $\{y_m^{(i)}\}_{m\in\mathcal{C}_n^{(i)}}$ can capture the $n$th signal component of $\mathbf{s}^{(i)}$. 
%To ensure every signal component in $\mathbf{s}^{(i)}$ is sensed at least once during the sampling process \eqref{eq:signal_vector_model}, it is reasonable to assume that $\mathcal{C}_n^{(i)}$ is nonempty (thus, $1\leq|\mathcal{C}_n^{(i)}|<M_i$). Moreover,  for each $1\leq i\leq I$, the non-zero entries of $\Phib^{(i)}$ %and the bias values $b_m^{(i)}$'s 
%are assumed to be independently drawn from continuous probability distributions. 

%(or, equivalently $\mathbf{D}^{(i)}$ and $\mathbf{s}$)
Through cooperation among local nodes, the objective for each node is to achieve perfect/stable recovery of its observation signal $\mathbf{s}^{(i)}$ in the presence of sparse noise, even when the local measurement size $M_i$ is limited. It is worth noting from \eqref{eq:partial_view_model} that recovering $\mathbf{s}^{(i)}$ is equivalent to simultaneously estimating the global signal $\mathbf{s}$ and the corresponding masking matrix $\mathbf{D}^{(i)}$. Therefore, the considered distributed signal recovery problem can be equivalently formulated as  %This can be done, according to \eqref{eq:partial_view_model}, by simultaneously solving the \textit{estimation} problem of recovering the global signal vector $\mathbf{s}$ as well as the \textit{identification} problem of recovering the local binary masking matrices $\mathbf{D}^{(i)}$'s. 
\begin{subequations}\label{eq:optimal_L0_overall_recovery}
\begin{align}
    \min_{\substack{\mathbf{s}},\{\mathbf{D}^{(i)}\}_{i=1}^I} & ~~\sum_{i=1}^I\left\|\mathbf{y}^{(i)}-\big|\Phib^{(i)}\mathbf{s}^{(i)}\big|^2\right\|_0 + \lambda\|\mathbf{s}\|_0 \label{eq:optimal_global_L0recovery11}\\  \text{subject to} & \quad \mathbf{s}^{(i)}= \mathbf{D}^{(i)}\mathbf{s}, \quad 1\leq i\leq I.\\ 
     & \quad d_{n}^{(i)}\in\{0,1\}, \quad 1\leq n\leq N, 1\leq i\leq I, %\forall n\in\mathcal{T},%~ 1\leq i\leq I,
\end{align}\label{eq:optimal_L0}
\end{subequations}
where $\lambda>0$. Instead of solving the above $\ell_0$-minimization problem %in \eqref{eq:optimal_L0} 
directly, we propose a distributed signal reconstruction algorithm that improves local recovery by enabling inter-node collaboration without explicitly sharing each device’s measurement vector $\mathbf{y}^{(i)}$ with the central server. This architecture helps mitigate the risk of signal leakage under passive attacks such as eavesdropping \cite{Butun20}.  %Moreover, due to privacy considerations, we aim to design distributed signal reconstruction algorithms that allow local edge devices to jointly recover the global signal $\mathbf{s}$ without explicitly sharing individual magnitude-squared partial-view measurements $\mathbf{y}^{(i)}$ with the central server. In addition, 
%Also, the local sensing matrices $\Phib^{(i)}$'s are unknown on the central server. 
The proposed distributed multi-view signal retrieval scheme consists of two stages: \textit{global signal amplitude recovery} and \textit{ Nonzero local signal entry retrieval}, which will be introduced in Sections \ref{sec:noise_free_issue} and \ref{sec:masking_matrix_issue}, respectively. Our proposed method is based on the following combinatorial structures, known as disjunct matrices, which are commonly found in sparse sensing matrices. %Our proposed recovery scheme is based on the following combinatorial structures within sparse sensing matrices, known as disjunct matrices.  %Our proposed recovery scheme is based on basic combinatorial structures on sensing matrices, called disjunct matrices and defined below.
\begin{definition}
    \cite{Macula97} An $M\times N$ matrix $\Phib$ is said to be $K^t$-disjunct %where $t\geq1$,
    if, for each column of $\Phib$, there exist %at least 
    $t+1$ elements in its support that do not lie in the union of the supports of any other $K$ columns.
    
    %if the support of any column is not entirely contained within the union of the supports of any other $K$ columns, i.e., $\mathcal{C}_{n_0}\not\subset(\mathcal{C}_{n_1}\cup\mathcal{C}_{n_2}\cup\cdots\cup\mathcal{C}_{n_K})$ holds for any distinct $n_0,n_1,\dots,n_K\in\{1,\dots,N\}$, where $\mathcal{C}_{n}\subset\{1,\dots,M\}$ is the support set of the $n$th column in $\Phib$. %\cup_{n'\in\{n_1,\dots,n_K\}}$. 
\end{definition}
Disjunct matrices have been widely adopted in the literature on CS and group testing, e.g., \cite{Matsumoto24,Truong20,Rescigno24}. As shown in \cite{Truong20,Rescigno24}, such sensing matrices can be generated with $M=\mathcal{O}(Kt\log(N))$ using a Bernoulli random design with $q=1/(K+1)$. Although the sensing matrix $\Phib^{(i)}$ employed at each device may not strictly satisfy the disjunct property due to a limited number of measurements, this issue can be addressed by partitioning the sensing matrices $\Phib^{(i)}$'s into $B$ disjoint blocks $\Phib_1$, $\Phib_2$, $\dots$, $\Phib_B$ (each of approximately equal size). %As a result, Assumption \ref{ass:group_wise_disjunct} is adopted throughout the paper and is considered reasonable as long as the number of devices is sufficiently large. Additionally,   
As a result, the following assumptions are applied %to each block-wise sensing matrix
throughout the paper. Assumption \ref{ass:group_wise_disjunct} is considered reasonable as long as the number of devices is large enough to provide sufficient measurements in each group. The condition is particularly well-suited to large-scale sensor networks. % , especially in large-scale sensor networks, provided that the number of devices is sufficiently large to ensure enough measurements in each group.} %as long as the number of devices is large enough to provide sufficient measurements in each group. }%, provided that the number of devices is sufficiently large to guarantee an adequate number of measurements per group.} %Assumption \ref{ass:group_wise_disjunct} is considered reasonable as long as the number of devices is sufficiently large. %As a result, the following assumption holds for each block-wise sensing matrix, provided that the number $I$ of devices is sufficiently large. %at each device may not satisfy disjunct property due to small number of measurements, we can address this by partitioning the sensing matrices $\Phib^{(i)}$'s into $B$ disjoint bloks $\Phib_1$, $\Phib_2$, $\dots$, $\Phib_B$ of roughly equal size. Hence, the following assumption is satisfied for each block sensing matrices as long as the number $I$ of devices is large enough. The following assumptions are made in the sequel.
\begin{assumption}\label{ass:group_wise_disjunct}
    Partition the sensing matrices $\{\Phib^{(i)}\}_{i=1}^I$ into $B$ disjoint groups, denoted by $\Phib_1$, $\dots$, $\Phib_B$, each of size $M'_b\times N$, where $M'_b=\sum_{i\in\mathcal{I}_b}M_i$, and $\mathcal{I}_b$ denotes the index set of devices associated with $\Phib_b$. 
    For each $1\leq b\leq B$, the group-wise sensing matrix $\Phib_b$ satisfies the $K^t$-disjunct property.
\end{assumption}
\begin{assumption}\label{ass:group_wise_model}
    %Let $\mathcal{I}_b$ be the $b$th group of devices corresponding to $\Phib_b$ as defined in Assumption 1, where $1\leq b\leq B$. We assume that this 
    The partition, i.e., $\mathcal{I}_1, \mathcal{I}_2, \dots,\mathcal{I}_B$, is known at the central server.
\end{assumption}

\section{Cooperative Global Sparse Signal Amplitude Reconstruction }\label{sec:noise_free_issue}

In this section, we address the global signal amplitude reconstruction problem using the models in \eqref{eq:partial_view_model} and \eqref{eq:signal_vector_model} through a distributed approach. %the problem of distributed phase retrieval from \eqref{eq:signal_vector_model} in the noise-free case, where $\mathbf{w}^{(i)}=\mathbf{0}$ for all $i$. Then, we will extend the results to the noisy case in Section \ref{sec:noise_corruption_issue}. 
The proposed distributed recovery algorithm consists of two stages: \textit{global signal support identification} and \textit{ nonzero entry amplitude recovery}, which will be shown in Sections \ref{sec:noisefree_support_est} and \ref{sec:signal_recovery}, respectively. 

\subsection{Global Signal Support Identification }\label{sec:noisefree_support_est}
We first note from \eqref{eq:signal_vector_model} that the compressed signal $\Phib^{(i)}\mathbf{s}^{(i)}$, $1\leq i\leq I$, is in general sparse due to the sparse nature of both the local observable signal $\mathbf{s}^{(i)}$ and sensing matrix $\Phib^{(i)}$. Consequently, sparse noise corruption results in a large number of zero entries in $\mathbf{y}^{(i)}$. %As a result, under sparse noise corruption, a large number of entries in $\mathbf{y}^{(i)}$ are zero.  %since they do not capture any nonzero signal entry, i.e., $\underline{\boldsymbol{\phi}}_{m}^{(i)}\mathbf{s}^{(i)}=0$.  
Specifically, for $n\in\mathcal{K}$, if the $n$th signal component is observable at edge node $i$, i.e., $d_n^{(i)}=1$, then the measurements indexed by $\mathcal{C}_n^{(i)}$ are allowed to have $y_m^{(i)}=|\underline{\boldsymbol{\phi}}_{m}^{(i)}\mathbf{s}^{(i)}|^2\neq0$ in general,  since $\underline{\boldsymbol{\phi}}_{m}^{(i)}$ is nonzero in its $n$th element, and so is ${\bf s}^{(i)}$. However, this may not hold for $n\notin\mathcal{K}$. In particular, under Assumption \ref{ass:group_wise_disjunct} and in the absence of noise, %in the noise-free setting,
each group is guaranteed to contain at least $t+1$ zero-valued measurements for every index $n\notin\mathcal{K}$. %at least $t+1$ zero-valued measurements exist for each index $n\notin\mathcal{T}$. %In particular, according to Assumption 1, there exist at least $t+1$ zero-valued measurements associated with indices $n\notin\mathcal{T}$. 
Motivated by this observation, we develop a two-step group-wise cooperative protocol for support identification, as follows. %propose the following two-step counting-based cooperative support identification protocol. %consists of two steps: local partial support inference and fusion-based support identification, as summarized below.
\vspace{.05cm}

\textbf{Step I: Local Partial Support Inference}
\begin{itemize}
    \item Based on the local measurement vector $\mathbf{y}^{(i)}$, the $i$th edge device first computes, for each index $n\in\{1,\ldots,N\}$, the number of measurements in $\{y_m^{(i)}\}_{m\in\mathcal{C}_n^{(i)}}$ that indicate $n$ is inactive, namely, $u_n^{(i)}=\big|\big\{m\in\mathcal{C}_n^{(i)}\big| y_m^{(i)}=0\big\}\big|$. %which $n$ is inactive during the sparse sensing process, namely, %the ``average relative frequency'' that $n$ is activated during the sparse sensing process, namely,
    %\begin{align}\label{eq:local_count}
        %u_n^{(i)}=%\frac{1}{|\mathcal{C}_n^{(i)}|}
        %\sum_{m\in\mathcal{C}_n^{(i)}}1\left\{y_m^{(i)}=0\right\}
        %\left|\big\{m\in\mathcal{C}_n^{(i)}\big| y_m^{(i)}=0\big\}\right|.
    %\end{align}
    %where $1\{\cdot\}$ is the indicator function.
    \item Afterwards, device $i$ stacks all $u_n^{(i)}$'s to get a vector $\mathbf{u}^{(i)}=[u_1^{(i)}\dots u_N^{(i)}]^{T}$, and then forwards it to the central server. 
\end{itemize}\vspace{.1cm}

\textbf{Step II: Fusion for Global Support Identification}
\begin{itemize}
    \item Upon receiving the vectors $\mathbf{u}^{(i)}$'s from local devices, the central server identifies the global signal support using the following data aggregation mechanism, which is based on a simple group-wise counting rule: %according to the following data aggregation mechanism which is built on a simple group-wise counting rule:
    \begin{align}\label{eq:global_support_est}
    \hat{\mathcal{K}}=\bigg\{ 1\leq n\leq N %\{1,\dots,N\}
    \bigg|~\exists 1\leq b\leq B, %\in\{1,\dots,B\},
    ~\mbox{s.t.}~\sum_{i\in\mathcal{I}_b}u_n^{(i)}<\eta\bigg\}, \end{align} 
    where $\mathcal{I}_b$ is defined in Assumption \ref{ass:group_wise_model} and $\eta>0$ represents a decision threshold.
    \item To conserve energy resources, the estimated signal support $\hat{\mathcal{K}}$ is not directly transmitted back to every local device for collaboratively recovering the amplitudes of nonzero signal entries, as will be shown later. Instead, the central server transmits each index in $\hat{\mathcal{K}}$ only to certain groups of devices that satisfy the constraint in \eqref{eq:global_support_est}.
    %\item The estimated signal support $\hat{\mathcal{T}}$ is then sent back to each local device.
\end{itemize}
%is that the local measurements $\mathbf{y}^{(i)}$'s in \eqref{eq:signal_vector_model} are in general sparse due to the sparse nature of both the signal vectors $\mathbf{s}^{(i)}$'s and sensing matrices $\Phib^{(i)}$'s.

It is worthwhile to note that the proposed protocol enables local nodes to collaboratively identify the global signal support while keeping both their individual measurement data $\mathbf{y}^{(i)}$ and sensing matrix $\Phib^{(i)}$ on-device, thereby reducing the risk of eavesdropping in the WSN.  %, thereby mitigating privacy concerns. {\blue Such a decentralized architecture helps preserve data privacy by ensuring that neither raw measurements nor sensing matrices are shared with the server, thereby mitigating the risk of signal leakage under passive attacks such as eavesdropping \cite{Butun20}.} 
To establish a mathematical performance guarantee for the proposed identification scheme, we define $\mathbf{w}_b\in\mathbb{R}^{M'_b}$, $1\leq b\leq B$, as the \textit{group-wise} sparse noise (outliers) in group $b$. Specifically, $\mathbf{w}_b$ consists of the noise vectors $\{\mathbf{w}^{(i)}\}_{i\in\mathcal{I}_b}$. %Toward a mathematical performance guarantee for the proposed scheme, for each $1\leq b\leq B$, we define $\mathbf{w}_b\in\mathbb{R}^{|\mathcal{I}_b|}$ as the group-wise sparse noise in the group $b$.
Below, we show that, under quite mild conditions, the proposed distributed cooperative support estimate in \eqref{eq:global_support_est} is guaranteed to be exact, i.e., $\hat{\mathcal{K}}=\mathcal{K}$. %To proceed, let us first 
%In the following, we provide a sufficient condition that guarantees the equivalence between the consistency on anchornode observations in (15) and the consistency of sparse representations in (14c). Our sufficient condition is characterized in terms of the restricted isometric property (RIP).
\begin{theorem}\label{th:exact_support_estimate}
Consider the signal model in \eqref{eq:signal_vector_model} and let $K_o={\displaystyle \max_{1\leq b\leq B}}\|\mathbf{w}_b\|_0$ and $\alpha={\displaystyle \max_{1\leq n\leq N}}\sum_{i=1}^I(1-d_n^{(i)})$. Under Assumption \ref{ass:group_wise_disjunct}, if $K_o<t/2$ and $\alpha\leq B-1$, then $\hat{\mathcal{K}}$ in \eqref{eq:global_support_est} with $\eta\in (K_o,t+1-K_o)$ exactly recovers the global signal support $\mathcal{K}$. %and let $\alpha={\displaystyle \max_{1\leq n\leq N}}\sum_{i=1}^I(1-d_n^{(i)})$. Then we partition all $I$ edge devices, indexed by the set $\mathcal{I}=\{1,\dots,I\}$, into $\alpha+1$ groups, that is, $\bigcup_{j=1}^{\alpha+1}\mathcal{I}_j=\mathcal{I}$, where $\mathcal{I}_1,\dots,\mathcal{I}_{\alpha+1}$ are disjoint. Let $\beta={\displaystyle \min_{1\leq j\leq \alpha+1, 1\leq n\leq N}}\sum_{i'\in\mathcal{I}_j}|\mathcal{C}_n^{(i')}|$ and $\gamma={\displaystyle \max_{1\leq j\leq \alpha+1}\max_{1\leq n\neq n'\leq N}}\sum_{i'\in\mathcal{I}_j}|\mathcal{C}_n^{(i')}\cap\mathcal{C}_{n'}^{(i')}|$.  Then, when $\frac{\beta+\gamma-2}{(\alpha+2)\gamma}>K$, the support estimate $\hat{\mathcal{T}}$ (defined in \eqref{eq:global_support_est}) with  $\eta_n\in \big[(\alpha+1)K\gamma/C_n, (\beta-K\gamma+\gamma-2)/C_n\big)$ exactly identifies the signal support $\mathcal{T}$. %[later on]
\end{theorem}
\begin{IEEEproof}
See Appendix \ref{app.A}.
%Later on. 
\end{IEEEproof}

Theorem \ref{th:exact_support_estimate} shows that, based on Assumption \ref{ass:group_wise_disjunct}, exact support identification can be achieved via the group-wise counting rule in \eqref{eq:global_support_est} when the blockage level of each signal component and the number of outliers in the network are sufficiently small. %and the column support of the effective sensing matrix for each group is large enough to capture the nonzero signal entries successfully. Indeed, it can be shown that if the sensing matrix $\Phib^{(i)}$ satisfies the UFF family [??] for all $i$ with parameter (?,?,?,?), then for $K>1$, the inequality $\frac{\beta+\gamma-1}{\gamma(\alpha+2)}>K+\frac{K_o}{\gamma}$ holds by letting $\zeta=\frac{1}{2(K+K_o/\gamma)}$.

%\textcolor{red}{[Ming: It is worthwhile to note that, the theoretical result about perfect support identification proposed in our TIT paper is the special case of Theorem \ref{th:exact_support_estimate}, where $I=1$ and $\alpha=0$.] }

%The above theorem shows that perfect support identification can be achieved by means of a simple counting rule in \eqref{eq:support_estimation4} if the signal support is small enough and the column support of the sensing matrix is large enough to successfully capture the nonzero signal entries. In fact,  for $K\geq 2$, the sparsity condition is satisfied with an $(N,M,d,1,\frac{1}{2K})$-UFF family of matrices since, by setting $\frac{r+1}{d}=\frac{1}{2K}<\frac{1}{2K-1}<\frac{d+2K-3}{(2K-1)d}$, it follows that $\frac{d+r-2}{2r}>K$. Notably, as shown in \cite{LFlodin19}, an $(N,M,d,1,\frac{1}{2K})$-UFF family of matrices can be  generated using error-correcting codes with $M=\mathcal{O}(K^2\log N)$ and $d=\mathcal{O}(K\log N)$. 

\subsection{Collaborative Global Signal Amplitude Recovery}
%\subsection{Distributed Global Signal Recovery}
%\subsection{Nonzero Entry Recovery of Global Signal}
% Entry Retrieval}
\label{sec:signal_recovery}
Assuming perfect support estimation $\hat{\mathcal{K}}=\mathcal{K}$, we go on to address the issue of recovering $|s_n|$ for all $n\in\mathcal{K}$. As mentioned before, only the groups of local devices satisfying $\sum_{i\in\mathcal{I}_b}u_n^{(i)}<\eta$ %for some $n\in\mathcal{T}$ 
will receive support knowledge from the central server, in order to reduce energy consumption. Specifically, %for each $n\in\mathcal{T}$, 
let $b_n\in\Bar{\mathcal{I}}_n\triangleq\{1\leq b\leq B|\sum_{i\in\mathcal{I}_b}u_n^{(i)}<\eta\}$ be the index of the corresponding group that meets the constraint in \eqref{eq:global_support_est}. %To proceed, for $n\in\mathcal{T}$, we denote by $\mathcal{C}_{n,b}\subset\{1,\dots,M'\}$ the support set of column $n$ in $\Phib_b$, where $b\in\{1,\dots,B\}$ is the group index that satisfies the constraint in \eqref{eq:global_support_est}.   After obtaining perfect support estimation (i.e., $\hat{\mathcal{T}}=\mathcal{T}$) from the central server, each local device is then ready to collaboratively recover the nonzero entries $\{s_n\}_{n\in\mathcal{T}}$ of the global signal $\mathbf{s}$. %Let us first consider the noise-free case (i.e., $\mathbf{w}^{(i)}=\mathbf{0}$),
%For $n\in\mathcal{T}$, 
By definition of $\mathcal{C}_n^{(i)}$, we can observe from \eqref{eq:signal_vector_model} that for each $i\in\mathcal{I}_{b_n}$, 
the $m$th entry of $\mathbf{y}^{(i)}$ at device $i$, i.e., $y_m^{(i)}$, captures the $n$th  entry of the local signal $\mathbf{s}^{(i)}$ only if $m\in\mathcal{C}_n^{(i)}$. %and $d_n^{(i)}=1$. 
In particular, %based on \eqref{eq:partial_view_model}, 
when $m\in\Bar{\mathcal{C}}_n^{(i)}\triangleq\mathcal{C}_n^{(i)}\setminus\bigcup_{n'\in\mathcal{K}\setminus\{n\}}\mathcal{C}^{(i)}_{n'}$, %and $d_n^{(i)}=1$, %where $\Bar{\mathcal{C}}_n^{(i)}\triangleq\mathcal{C}_n^{(i)}\setminus\bigcup_{n'\in\mathcal{T}\setminus\{n\}}\mathcal{C}_{n'}$,
the local measurement $y_m^{(i)}$ reduces to the following form: %captures only the signal component $s_n$, which can be expressed as 
\begin{align}\label{eq:nonzero_sol}
     y_{m}^{(i)}%=\left|\phi_{m,n}^{(i)}d_n^{(i)}s_n+b_{m}^{(i)}\right|^2+w_m^{(i)}
     =\left|\phi_{m,n}^{(i)}d_n^{(i)}s_n\right|^2+w_m^{(i)}, ~m\in\Bar{\mathcal{C}}_n^{(i)},~i\in\mathcal{I}_{b_n}.
\end{align}
Therefore, with some manipulations, the amplitude of $s_n$ can be expressed as $|s_n|=\frac{\sqrt{y_{m}^{(i)}}}{|\phi_{m,n}^{(i)}|}$ in the full-view noiseless case (i.e., $\mathbf{D}^{(i)}=\mathbf{I}_N$ and $\mathbf{w}^{(i)}=\mathbf{0}$ in \eqref{eq:signal_vector_model} for all $i$).  Note that $\Bar{\mathcal{C}}_n^{(i)}$ may be empty due to the limited number of measurements available at local device $i$. However, when the number of devices ($I$) is sufficiently large for Assumption \ref{ass:group_wise_disjunct} to hold, %by Assumption \ref{ass:group_wise_disjunct}, 
at least $t+1$ measurements in the device group $\mathcal{I}_{b_n}$ are guaranteed to satisfy the expression in \eqref{eq:nonzero_sol}. Thus, in the full-view noiseless case, perfect global signal amplitude recovery can be easily achieved by having these devices transmit their computed ratios $\frac{\sqrt{y_{m}^{(i)}}}{|\phi_{m,n}^{(i)}|}$ to the central server. 

In the presence of noise, the equation $|s_n|=\frac{\sqrt{y_{m}^{(i)}}}{|\phi_{m,n}^{(i)}|}$ may no longer hold, even in the full-view scenario. However, since $\mathbf{w}^{(i)}$ is sparse, many $w_m^{(i)}$ in \eqref{eq:nonzero_sol} are zero, implying that this equation %about $|s_n|$
may still hold for the majority of devices in $\mathcal{I}_{b_n}$ whose measurements follow the expression in \eqref{eq:nonzero_sol}. Motivated by this observation, %upon receiving these computed ratios $\frac{\sqrt{y_{m}^{(i)}}}{|\phi_{m,n}^{(i)}|}$, 
the central server estimates $|s_n|$ based on the following majority rule, which utilizes the ratios $\frac{\sqrt{y_{m}^{(i)}}}{|\phi_{m,n}^{(i)}|}$, instead of the raw data $y_m^{(i)}$, received from %certain 
local devices:
\begin{align}\label{eq:nonzero_amp_sol_1}
     |\hat{s}_n|=\arg\max_{s\in\bigcup_{ b_n\in\Bar{\mathcal{I}}_n}\mathcal{F}_{b_n}}\sum_{b_n\in\Bar{\mathcal{I}}_n}\sum_{s'\in\mathcal{F}_{b_n}}1\left\{s=s'\right\},
\end{align}
where $\mathcal{F}_{b_n}=\big\{\frac{\sqrt{y_{m}^{(i)}}}{|\phi_{m,n}^{(i)}|}|i\in\mathcal{I}_{b_n}, m\in\Bar{\mathcal{C}}_n^{(i)} \big\}$ and $1\{\cdot\}$ is the indicator function. The following theorem provides a sufficient condition that guarantees exact signal amplitude recovery using \eqref{eq:nonzero_amp_sol_1} in the partial-view scenario with sparse noise corruption. %under sparse noise corruption in the partial-view scenario. shows that In the presence of noise, Under the same conditions as made in Theorem \ref{th:exact_support_estimate} so that $\hat{\mathcal{T}}=\mathcal{T}$,  %As a result, perfect global signal amplitude recovery can be done in the full-view noiseless case by those devices transmitting their ratios $\frac{\pm\sqrt{y_{m}^{(i)}}-b_{m}^{(i)}}{\phi_{m,n}^{(i)}}$ to the central server.   

\begin{theorem}\label{th:recovery_sparse_noise}
Under the same setting as in Theorem \ref{th:exact_support_estimate} (hence, $\hat{\mathcal{K}}=\mathcal{K}$), the central server exactly recovers the amplitude of the $K$-sparse global signal $\mathbf{s}$ using the proposed estimate of $|s_n|$ in \eqref{eq:nonzero_amp_sol_1}.%, under the multi-view measurement model \eqref{eq:signal_vector_model} and $K_o$-sparse noise corruption. 
\end{theorem}
\begin{IEEEproof}
Appendix \ref{app.B}.
\end{IEEEproof}
After obtaining the amplitude of the global signal, the central server sends it back to local devices to facilitate the reconstruction of their respective local signals, as discussed in the following section. 

%\section{Local Masking Matrix Identification }\label{sec:masking_matrix_issue}
\section{Local Signal Reconstruction }\label{sec:masking_matrix_issue}

With perfect amplitude reconstruction of $\mathbf{s}$ (i.e., $|\hat{\mathbf{s}}|=|\mathbf{s}|$), each edge device then individually addresses the problem of recovering its local signal $\mathbf{s}^{(i)}$. Let us rewrite the global signal $\mathbf{s}$ as  %which can be done by solving the following optimization problem:
%First, we express $\mathbf{s}$ in polar form as: 
\begin{align}\label{eq:signal_polar_form}
    \mathbf{s}=\text{diag}\left(|s_{1}|,|s_{2}|,\dots,|s_{N}|\right)\mathbf{p}, 
\end{align}
where $\mathbf{p}=[p_1\cdots p_N]^T\in\mathbb{R}^N$ is the phase vector of $\mathbf{s}$ with entries $p_n=\text{sign}(s_n)$ for all $n$. Then based on \eqref{eq:partial_view_model} and \eqref{eq:signal_polar_form}, the local signal can be further expressed as
\begin{align}\label{eq:signal_polar_form2}
    \mathbf{s}^{(i)}=\mathbf{D}^{(i)}\mathbf{s}=\text{diag}\left(|s_{1}|,|s_{2}|,\dots,|s_{N}|\right)\mathbf{h}^{(i)}, %\underbrace{\mathbf{D}^{(i)}\mathbf{p}}_{\triangleq\mathbf{h}}, 
\end{align}
where $\mathbf{h}^{(i)}\triangleq\mathbf{D}^{(i)}\mathbf{p}=[d_1^{(i)}p_1~ d_2^{(i)}p_2 \dots d_N^{(i)}p_N]^T$. Hence, it can be seen from \eqref{eq:signal_polar_form2} that under perfect amplitude retrieval, the problem of recovering $\mathbf{s}^{(i)}$ reduces to determining $\mathbf{h}^{(i)}$. %finding $\mathbf{s}^{(i)}$ is equivalent to finding $\mathbf{h}^{(i)}$.
With the aid of the estimated signal support and \eqref{eq:signal_polar_form2}, the local measurement model in \eqref{eq:signal_vector_model} admits the following expression:
\begin{align}\label{eq:signal_vector_model2}
    \mathbf{y}^{(i)}%&=\left|\Phib^{(i)}_{\mathcal{T}}\mathbf{s}^{(i)}_{\mathcal{T}}\right|^2+\mathbf{w}^{(i)}\notag\\
    =\bigg|\underbrace{\Phib^{(i)}_{\mathcal{K}}\text{diag}\left(|s_{n_1}|,|s_{n_2}|,\dots,|s_{n_K}|\right)}_{\triangleq\Tilde{\Phib}^{(i)}_{\mathcal{K}}}\mathbf{h}^{(i)}_{\mathcal{K}}\bigg|^2+\mathbf{w}^{(i)}, 
\end{align}
where $\mathcal{K}=\{n_1,\dots,n_K\}$, $\Phib^{(i)}_{\mathcal{K}}\in\mathbb{R}^{M_i\times K}$ is constructed by selecting the columns of $\Phib^{(i)}$ indexed by $\mathcal{K}$, and  $\mathbf{h}^{(i)}_{\mathcal{K}}\in\mathbb{R}^{K}$ is obtained by retaining the entries of $\mathbf{h}^{(i)}$ indexed by $\mathcal{K}$. %Therefore, based on
With \eqref{eq:signal_vector_model2}, %\eqref{eq:signal_polar_form2} and \eqref{eq:signal_vector_model2},   %Let $\mathbf{D}^{(i)}_{\mathcal{T}}=\text{diag}(d^{(i)}_{n_1}\cdots d^{(i)}_{n_K})\in\mathbb{C}^{K\times K}$ be the matrix obtained by retaining the diagonal entries of $\mathbf{D}^{(i)}$ indexed by $\mathcal{T}$. Then, using \eqref{eq:partial_view_model} and \eqref{eq:signal_vector_model2}, 
the optimal solution for $\mathbf{s}^{(i)}$ can be found by solving the following $\ell_0$-minimization problem
\begin{subequations}\label{eq:optimal_local_recovery}
\begin{align}
    \min_{\substack{\mathbf{x}}=[x_1\cdots x_K]^T\in\mathbb{R}^K} & ~~\left\|\mathbf{y}^{(i)}-\big|\Tilde{\Phib}^{(i)}_{\mathcal{K}}\mathbf{x}\big|^2\right\|_0 \label{eq:optimal_local_recovery11}\\  \text{subject to} & \quad x_{j}\in\{0,\pm1\}, \quad 1\leq j\leq K. %\forall n\in\mathcal{T},%~ 1\leq i\leq I,
\end{align}
\end{subequations}
%where %$\mathbf{D}^{(i)}_{\mathcal{T}}\in\mathbb{C}^{K\times K}$ is obtained by retaining the diagonal entries of $\mathbf{D}^{(i)}$ indexed by $\mathcal{T}$, and 
%$\mathbf{s}_{\mathcal{T}}=[s_{n_1}\cdots s_{n_K}]$ is the subvector of $\mathbf{s}$ obtained by retaining its nonzero entries. The following theorem shows that, under mild conditions,  solving the above optimization problem ensures exact local signal recovery (up to a global sign ambiguity) in the noiseless case. We consider a bipartite graph $G_i$ consisting of $N$ left nodes and $M_i$ right nodes, where the edges are determined by the local sensing matrix $\Phib^{(i)}$ such that left node $n$ is connected to right node $m$ if and only if $\phi_{m,n}^{(i)}\neq0$. %exact local signal recovery can be achieved by solving the above optimization problem in the noise case.
%To establish a theoretical performance guarantee, 
Towards an analytic performance guarantee for the above recovery scheme, we define a bipartite graph $G_i$ associated with the local sensing matrix $\Phib^{(i)}$. This graph consists of $N$ left nodes and $M_i$ right nodes, where an edge exists between left node $n$ and right node $m$ if and only if $\phi_{m,n}^{(i)}\neq0$ in  $\Phib^{(i)}$. %Based on this graph structure, 
The following theorem demonstrates that, under a mild condition on the graph structure, solving the above optimization problem achieves exact local signal recovery (up to a global sign ambiguity) in the noiseless case. 
\begin{theorem}\label{th:recovery_local_signal}
Given perfect global signal amplitude recovery and a connected graph $G_i$, the local signal $\mathbf{s}^{(i)}$ can be exactly recovered with probability one by solving the noiseless $\ell_0$-norm minimization problem in \eqref{eq:optimal_local_recovery}. %If the number of outliers is less than $\frac{M_i}{2}$, the local signal $\mathbf{s}^{(i)}$ can be exactly recovered with probability one via solving noiseless $\ell_0$-minimization problem in \eqref{eq:optimal_local_recovery}.  %By assuming that the nonzero entries of $\Phib^{(i)}$ and the bias values are continuous random variables, the uniqueness of the optimization problem in \eqref{eq:optimal_local_recovery} can be ensured, provided that the number of outliers is less than $\frac{M_i}{2}$.
\end{theorem}
%\begin{theorem}\label{th:recovery_local_signal2}
%Consider a bipartite graph $G_i$ with $N$ left nodes and $M_i$ right nodes, where an edge exists between left node $n$ and right node $m$ if and only if $\phi_{m,n}^{(i)}\neq0$ in $\Phib^{(i)}$. Assume perfect global signal recovery, and that the nonzero entries of $\Phib^{(i)}$ are continuous random variables. If the number of outliers is less than $\frac{M_i}{2}$, the local signal $\mathbf{s}^{(i)}$ can be exactly recovered with probability one via solving the optimization problem in \eqref{eq:optimal_local_recovery}.  %By assuming that the nonzero entries of $\Phib^{(i)}$ and the bias values are continuous random variables, the uniqueness of the optimization problem in \eqref{eq:optimal_local_recovery} can be ensured, provided that the number of outliers is less than $\frac{M_i}{2}$.
%\end{theorem}
\begin{IEEEproof}
See Appendix \ref{app.C}.
%Let $\tilde{\phi}_{m,n}^{(i)}$ be the $(m,n)$th entry of $\Tilde{\Phib}^{(i)}_{\mathcal{T}}$. Then, under the assumption that the nonzero entries of $\mathbf{s}$ and $\Phib^{(i)}$ are continuous random variables, $\tilde{\phi}_{m,n}^{(i)}$ is also drawn from a continuous probability distribution. 
    %Let $\hat{\mathbf{D}}^{(i)}_{\mathcal{T}}$ be the solution to \eqref{eq:optimal_local_recovery}. Then under the continuous random variables assumption, it can be directly deduced from \eqref{eq:partial_view_model} and \eqref{eq:signal_vector_model} that in the noise-free case, the value of the objection function in \eqref{eq:optimal_local_recovery11} is 0 ($M_i$, respectively) with probability one, if $\hat{\mathbf{D}}^{(i)}_{\mathcal{T}}\neq\mathbf{D}^{(i)}_{\mathcal{T}}$ ($\hat{\mathbf{D}}^{(i)}_{\mathcal{T}}=\mathbf{D}^{(i)}_{\mathcal{T}}$, respectively). The assertion of Theorem \ref{th:recovery_local_signal} is thus proved due to $\|\mathbf{w}^{(i)}\|_0<M_i/2$. %immediately follows since $\|\mathbf{w}^{(i)}\|_0<\frac{M_i}{2}$. 
\end{IEEEproof}
To avoid the NP-hardness and intractability of $\ell_0$-norm minimization, we adopt a widely-used $\ell_1$ norm relaxation approach in \eqref{eq:optimal_local_recovery}, leading to the following optimization problem
\begin{subequations}\label{eq:optimal_local_recovery_L1}
\begin{align}
    \min_{\substack{\mathbf{x}}=[x_1\cdots x_K]^T\in\mathbb{R}^K} & ~~\left\|\mathbf{y}^{(i)}-\big|\Tilde{\Phib}^{(i)}_{\mathcal{K}}\mathbf{x}\big|^2\right\|_1 \label{eq:optimal_local_recoveryL1_1}\\  \text{subject to} & \quad x_{j}\in\{0,\pm1\}, \quad 1\leq j\leq K. %\forall n\in\mathcal{T},%~ 1\leq i\leq I,
\end{align}
\end{subequations}
%or, equivalently,
%\begin{align}\label{eq:eq:optimal_local_recovery_L1_2}
    %\min_{\mathbf{d}^{(i)}=[d^{(i)}_{n_1}\cdots d^{(i)}_{n_K}]\in\{0,1\}^{K}} \left\|\mathbf{y}^{(i)}-\big|\Tilde{\Phib}^{(i)}_{\mathcal{T}}\mathbf{d}^{(i)}+\mathbf{b}^{(i)}\big|^2\right\|_1,
%\end{align}
%where $\{0,1\}^K$ represents the set of length-$K$ vectors with each entry being either 0 or 1, and $\Tilde{\Phib}^{(i)}_{\mathcal{T}}=\Phib^{(i)}_{\mathcal{T}}\text{diag}(s_{n_1},s_{n_2},\dots,s_{n_K})$. %with $s_{\mathcal{T},k}$ being the $k$th element of $\mathbf{s}_{\mathcal{T}}$, and $\mathbf{d}^{(i)}=[d^{(i)}_{n_1}\cdots d^{(i)}_{n_K}]$ with $n_k\in\mathcal{T}$. 
While solving the above ternary Boolean-constrained optimization problem typically requires an exhaustive search over $3^K$ possible combinations, it can be efficiently addressed with a two-stage projection approach that ensures tractable computational complexity. Specifically, we first solve the optimization problem in \eqref{eq:optimal_local_recovery_L1} without the ternary constraint using existing phase retrieval algorithms (e.g., robust-PhaseLift \cite{Yuanxin17}). Then, the obtained solution is projected onto the feasible region by rounding each entry to $-1$, $0$, or $1$. The simulation results demonstrate that the proposed two-stage projection approach can deliver comparable recovery performance to solving Problem \eqref{eq:optimal_local_recovery} directly.

%Hence, as long as $\hat{\mathcal{T}}=\mathcal{T}$, Theorem \ref{th:signal_recovery_motivate1} suggests the following simple signal recovery protocol: (i) for $n\in\mathcal{T}$, select three elements $m_{1},m_{2},m_{3}\in\Tilde{\mathcal{C}}_n$ and find the associated solution sets $\mathcal{F}^{(n)}_{m_1}$, $\mathcal{F}^{(n)}_{m_2}$ and $\mathcal{F}^{(n)}_{m_3}$ using \eqref{eq:sol_set1}; (ii) compute $s_n$ according to the formula \eqref{eq:nonzero_closed_form_re}.  The proposed signal recovery scheme is summarized in Algorithm \ref{alg::proposed_recovery_method}.

\begin{figure}[t]
\centering
\includegraphics[width=0.55\textwidth]{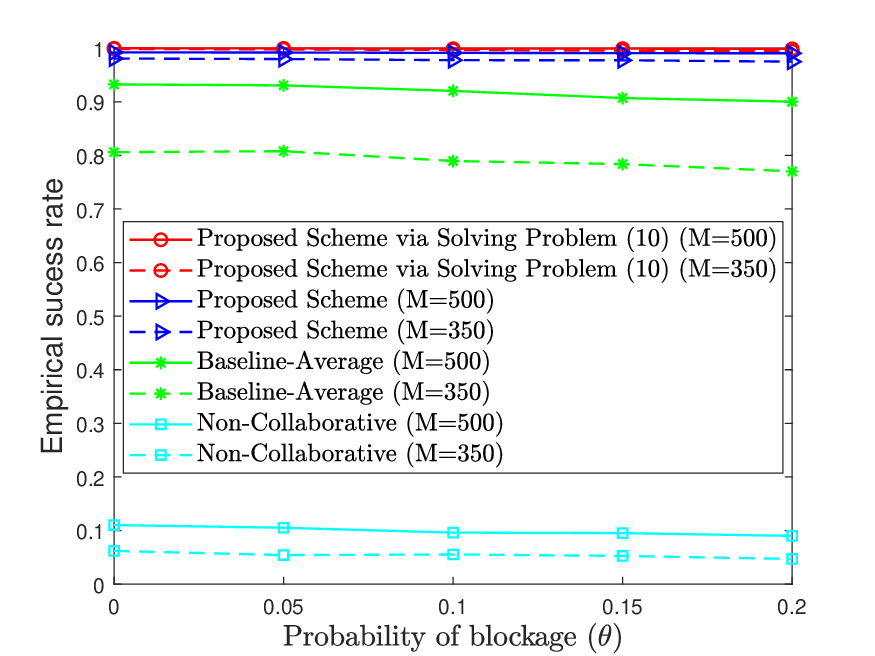}
\caption{{Performance comparisons of all methods for different values of $\theta$.}} %and the empirical solution.}
\label{Fig_SR_theta}
\vspace{-.3cm}
\end{figure}

\begin{figure}[t]
\centering
\includegraphics[width=0.55\textwidth]{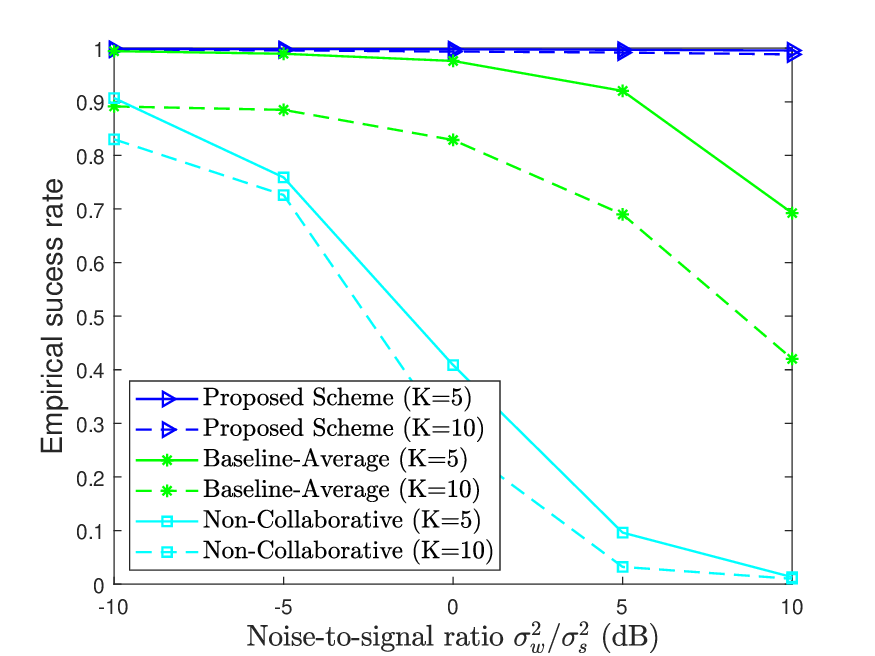}
\caption{{Performance comparisons of all methods for different values of NSR.}} %and the empirical solution.}
\label{Fig_SR_noise}
\vspace{-.3cm}
\end{figure}

%\vspace{-.5cm}
\section{Performance Evaluation}
%\section{Experimental Results}
\label{simulations}

In this section, computer simulations are provided to illustrate the effectiveness of the proposed cooperative signal recovery scheme in the multi-view sensing scenario. We consider a sensor network with $I=30$ devices and a global signal of dimension $N=2500$. %We consider the case where the dimension of the global signal is set to be $N=2500$ and 
The signal support $\mathcal{K}$ is selected uniformly at random, and the nonzero entries $s_n$, for $n\in\mathcal{K}$, are independently drawn from a standard Gaussian distribution, i.e., $\mathcal{N}(0,1)$. For simplicity of illustration, we assume that all devices have the same probability of partial-view observation, denoted by $\theta$. When this happens, a subset of the entries in $\mathbf{s}_{\mathcal{K}}$ is masked. Specifically, the number of masked entries (i.e., the blockage level) is randomly selected from a discrete uniform distribution over the interval $[1, K - 1]$. %When this happens, the blockage level is randomly selected from a discrete uniform distribution over the integer interval $[1,K-1]$. 
Furthermore, at each local device, each measurement is independently corrupted by an outlier with probability $0.05$, where the outlier is modeled as a random variable from $\mathcal{N}(0,\sigma_w^2)$. %consider a homogeneous sensing environment, wherein the partial-view probability is identical for all devices and is denoted by $\theta$. For simplicity of illustration, 
%For all $i$, the number of measurements per sensing device is fixed as $M_i=350$. The probability that a given measurement is corrupted by an outlier, modeled as a random variable from $\mathcal{N}(0,\sigma_w^2)$, is 0.05.
The sensing matrix $\Phib^{(i)}$ has a sparsity rate of $q=0.08$, and its nonzero entries are generated independently from $\mathcal{N}(0,1)$. The recovery performance of each algorithm is evaluated in terms of the success rate, which is defined as the ratio of the number of successful trials to a total of 500 independent runs across all devices. As in \cite{GWang18}, a trial is declared successful if the relative error of the local signal recovery is less than $10^{-3}$. %$\frac{\min(\|\hat{\mathbf{s}}-\mathbf{s}\|_2,\|\hat{\mathbf{s}}+\mathbf{s}\|_2)}{\|\mathbf{s}\|_2}$ is less than $10^{-3}$. The reconstruction quality of local signals is evaluated using the \textit{relative error} [??], defined as $GloRE\triangleq\min_{\omega\in[0,2\pi)}\|e^{i\omega}\hat{\mathbf{s}}-\mathbf{s}\|_2/\|\mathbf{s}\|_2$ for the global signal and $LocRE\triangleq\frac{1}{I}\sum_{i=1}^{I}\min_{\omega\in[0,2\pi)}\|e^{i\omega}\hat{\mathbf{s}}_i-\mathbf{s}_i\|_2/\|\mathbf{s}_i\|_2$ for the local signals. %In the discussions below,we compare with the following two baseline distributed sparse signal recovery methods:
We compare the proposed scheme with the following two baseline distributed sparse signal recovery methods: 1) \textbf{Non-Collaborative Recovery}, where each local device independently estimates its local signal $\mathbf{s}_i$ using the existing algorithm SPARTA \cite{GWang18}. 2) \textbf{Baseline-Average}, %which averages the non-collaborative estimates from all devices, following the IRAS approach proposed in [??]; then the central server broadcasts this global signal estimate to each device to update its nonzero signal entries. 
which computes the average of the non-collaborative estimates from all devices following the IRAS approach proposed in \cite{tian2023}, after which the central server broadcasts this global signal estimate to each device for updating its nonzero signal entries. %The simulation results are obtained from 500 independent trials.

We first compare the performance of the three methods with different probability of blockage $\theta$. For $K=5$, $B=10$ and NSR $\sigma_w^2/\sigma_s^2=\sigma_w^2=5$ dB, Fig. \ref{Fig_SR_theta} plots the empirical success rates of all methods with respect to $\theta$ when $M=350$ and $500$. We can see that our proposed scheme can achieve the highest recovery probability in all cases and outperform the two baselines, even when using fewer measurements (e.g., $M=350$ vs. $M=500$). This indicates that our proposed method is more efficient and stable than the competing solutions under partial-view blocking scenarios. Furthermore, the proposed projection-based approximation to Problem \eqref{eq:optimal_local_recovery_L1} exhibits comparable performance to that of the exact $\ell_0$-minimization in \eqref{eq:optimal_local_recovery}.  %Furthermore, the proposed projection-based approximation in \eqref{eq:optimal_local_recovery_L1} offers performance close to that of the exact $\ell_0$-minimization in \eqref{eq:optimal_local_recovery}. 
To further investigate the impact of noise level on signal reconstruction performance, we fix $M=500$ and $\theta=0.1$, and show in Fig. \ref{Fig_SR_noise} the empirical success rates of all methods for different values of NSR when $K=5$ and $K=10$. As the figure shows, although the performance of all methods degrades as NSR increases, the success rate of the proposed scheme decreases only slightly and remains above $0.95$ even when NSR exceeds 5 dB. This result, together with Fig. \ref{Fig_SR_theta}, confirms that the proposed distributed recovery scheme, which does not require sharing raw data, exhibits greater robustness to sparse noise (i.e., outlier) compared to the two baseline algorithms in multi-view blocking scenarios. %while the performances of all methods degrade as NSR increases, the success rate of the proposed scheme decreases only slightly and can provide a successful signal recovery with probability at least $0.95$ even in the high NSR. Moreover, since  The result again confirms that     
%To further investigate the impact of the noise level on signal reconstruction performance, for $M=500$ and $\theta=0.1$, Fig. ?? shows the empirical success rates of all methods w.r.t. different values of NSR, when $K=5$ and $10$.  Next, we examine the performance of all methods under sparse noise corruption. For $M=500$ and $\theta=0.1$, Fig. ??

%the proposed projection-based approximate approach can deliver comparable recovery performance to solving the $\ell_0$-minimization problem in \eqref{eq:optimal_local_recovery}. 
%yield better signal reconstruction performance even with fewer samples as compared to the two competitive solutions.

%The simulation results are obtained from 1000 independent trials. Following \cite{SGopi13}, the sparse sensing matrix of each device is randomly constructed using expanders (see Algorithm 1 in \cite{SGopi13}). The bias terms $b_m^{(i)}$, for $m=1,\dots,M_i$, are randomly drawn from a standard Gaussian distribution and are independent across sensor devices. and the network size is $I$.  We examine the performance of the proposed scheme for the image recovery task using the MNIST dataset \cite{YLeCun98}. 
%\vspace{-.5cm}

\appendices
\section{Proof of Theorem \ref{th:exact_support_estimate}} \label{app.A}

For $n\notin\mathcal{K}$, we first note that in the noiseless case and based on Assumption \ref{ass:group_wise_disjunct}, each group contains at least $t+1$ measurements with $y_m^{(i)}=0$ in the set $\{y_m^{(i)}\}_{m\in\mathcal{C}_n^{(i)},i\in\mathcal{I}_b}$. %Hence, we have
It then follows that $\sum_{i\in\mathcal{I}_b}u_n^{(i)}\geq t+1$ for all $b$.  Moreover, since the noise sparsity level is less than $K_o$, we have $\sum_{i\in\mathcal{I}_b}u_n^{(i)}\geq t+1-K_o$ for all $b$ under outlier corruption. Therefore, it follows immediately from \eqref{eq:global_support_est} that $n\notin\hat{\mathcal{K}}$, since $\eta<t+1-K_o$. This establishes $\hat{\mathcal{K}}\subseteq\mathcal{K}$. Next, we will prove $\mathcal{K}\subseteq\hat{\mathcal{K}}$. Let $n\in\mathcal{K}$. Since $\alpha\leq B-1$, there exists a group index $b'\in\{1,\dots,B\}$ such that the signal component $s_n$ is observable for all devices in the $b'$th group. Then by following similar arguments, it can be verified that $\sum_{i\in\mathcal{I}_{b'}}u_n^{(i)}\leq K_o<\eta$, which implies $n\in\hat{\mathcal{K}}$. Therefore, $\mathcal{K}\subseteq\hat{\mathcal{K}}$ and, thus, $\hat{\mathcal{K}}=\mathcal{K}$.  $\hfill\Box$  
%The proof is completed.    $\hfill\Box$

\section{Proof of Theorem \ref{th:recovery_sparse_noise}} \label{app.B}

As mentioned in Section \ref{sec:signal_recovery}, it follows from \eqref{eq:nonzero_sol} that for any $n\in\mathcal{K}$ and $m\in\Bar{\mathcal{C}}_n^{(i)}$, $\frac{\sqrt{y_{m}^{(i)}}}{|\phi_{m,n}^{(i)}|}=|s_n|$ is true for the case with $d_n^{(i)}=1$ and $w_m^{(i)}=0$. Moreover, since $\alpha\leq B-1$, we conclude that there exists a group index $b'\in\Bar{\mathcal{I}}_n$ such that all devices in group $b'$ have access to the $n$th nonzero entry $s_n$. In particular, by Assumption \ref{ass:group_wise_disjunct}, there are (at least) $t+1$ measurements in group $b'$ that admit the following expression: 
\begin{align}\label{eq:thm2_nonzero_sol}
     y_{m}^{(i)}%=\left|\phi_{m,n}^{(i)}d_n^{(i)}s_n+b_{m}^{(i)}\right|^2+w_m^{(i)}
     =\left|\phi_{m,n}^{(i)}s_n\right|^2+w_m^{(i)}. %~m\in\Bar{\mathcal{C}}_n^{(i)},~i\in\mathcal{I}_{b_n}.
\end{align}
Since $\|\mathbf{w}_{b'}\|_0\leq K_o$, %the support size of the group-wise noise $\mathbf{w}_{b'}$ is smaller than $K_o$,
\eqref{eq:thm2_nonzero_sol} asserts that there exist at least $t+1-K_o$ measurements in group $b'$ satisfying $\frac{\sqrt{y_{m}^{(i)}}}{|\phi_{m,n}^{(i)}|}=|s_n|$. Hence, $|s_n|$ belongs to the set $\bigcup_{ b_n\in\Bar{\mathcal{I}}_n}\mathcal{F}_{b_n}$, and the corresponding objective function value in \eqref{eq:nonzero_amp_sol_1} is bounded below by
\begin{align}\label{eq:thm2_nonzero_sol_2}
    \sum_{b_n\in\Bar{\mathcal{I}}_n}\sum_{s'\in\mathcal{F}_{b_n}}1\left\{|s_n|=s'\right\}\geq t+1-K_o>\frac{t}{2}+1.
\end{align}
where the last inequality holds due to $K_o<t/2$. Let $\Bar{s}_{n,1},\Bar{s}_{n,2}\in\bigcup_{ b_n\in\Bar{\mathcal{I}}_n}\mathcal{F}_{b_n}$ and $\Bar{s}_{n,j}\neq |s_n|$ for all $j=1,2$. Then $\Bar{s}_{n,1}$ and $\Bar{s}_{n,2}$ are almost surely distinct since  the nonzero entries of $\Phib^{(i)}$ are independently drawn from a continuous probability distribution. Consequently, it can be readily deduced that if $\Bar{s}_n\in\bigcup_{ b_n\in\Bar{\mathcal{I}}_n}\mathcal{F}_{b_n}$ and $\Bar{s}_n\neq|s_n|$, we have
\begin{align}\label{eq:thm2_nonzero_sol_3}
    \sum_{b_n\in\Bar{\mathcal{I}}_n}\sum_{s'\in\mathcal{F}_{b_n}}1\left\{\Bar{s}_n=s'\right\}\leq K_o+1<\frac{t}{2}+1.
\end{align}
The proof is thus completed. $\hfill\Box$  %\vspace{-.5cm}
%if $\Bar{s}_n\in\bigcup_{ b_n\in\Bar{\mathcal{I}}_n}\mathcal{F}_{b_n}$ and $\Bar{s}_n\neq|s_n|$.
%due to the fact that the nonzero entries of $\Phib^{(i)}$ are drawn from a continuous probability distribution.  %since the nonzero entries of $\Phib^{(i)}$ are drawn from continuous distribution.

\section{Proof of Theorem \ref{th:recovery_local_signal}} \label{app.C}

Let $\tilde{\phi}_{m,n}^{(i)}$ be the $(m,n)$th entry of $\Tilde{\Phib}^{(i)}_{\mathcal{K}}$. Then, under the assumption that the nonzero entries of $\mathbf{s}$ and $\Phib^{(i)}$ are continuous random variables, $\tilde{\phi}_{m,n}^{(i)}$ is also drawn from a continuous probability distribution. Let $\hat{\mathbf{x}}=[\hat{x}_1\cdots \hat{x}_K]^T$ be the solution to \eqref{eq:optimal_local_recovery}. Then, in the noiseless case, it follows that $\big\|\mathbf{y}^{(i)}-|\Tilde{\Phib}^{(i)}_{\mathcal{K}}\hat{\mathbf{x}}|^2\big\|_0=\big\|\mathbf{y}^{(i)}-|\Tilde{\Phib}^{(i)}_{\mathcal{K}}\mathbf{h}^{(i)}_{\mathcal{K}}|^2\big\|_0=0$, which in turn implies
%\begin{align}
    %\big|\Tilde{\Phib}^{(i)}_{\mathcal{T}}\hat{\mathbf{x}}\big|^2=\big|\Tilde{\Phib}^{(i)}_{\mathcal{T}}\mathbf{h}^{(i)}_{\mathcal{T}}\big|^2,
%\end{align}
%or equivalently,
\begin{align}\label{eq:thm3_nonzero_sol}
    \bigg| \sum_{n=1}^K\tilde{\phi}_{m,n}^{(i)}\hat{x}_n \bigg|=\bigg| \sum_{n=1}^K\tilde{\phi}_{m,n}^{(i)}h^{(i)}_{n} \bigg|, \quad 1\leq m\leq M_i,
\end{align}
where $h^{(i)}_{n}$ is the $n$th entry of $\mathbf{h}^{(i)}_{\mathcal{K}}$. %Based on \eqref{eq:thm3_nonzero_sol} 
With some manipulation, each equation in \eqref{eq:thm3_nonzero_sol} can be further expressed as
\begin{align}\label{eq:thm3_nonzero_sol2}
    \sum_{n=1}^K\tilde{\phi}_{m,n}^{(i)}(\hat{x}_n-h^{(i)}_{n})=0~\mbox{or}~ \sum_{n=1}^K\tilde{\phi}_{m,n}^{(i)}(\hat{x}_n+h^{(i)}_{n})=0. %\quad 1\leq m\leq M_i,
\end{align}
According to \eqref{eq:thm3_nonzero_sol2}, we have $|\hat{\mathbf{x}}|=|\mathbf{h}^{(i)}_{\mathcal{K}}|$ with probability one and, thus, the assertion of Theorem \ref{th:recovery_local_signal} follows directly from the fact that $G_i$ is connected. The proof is completed.  $\hfill\Box$
%$|\hat{\mathbf{x}}|=|\mathbf{h}^{(i)}_{\mathcal{T}}|$ holds. Furthermore, since $G_i$ is a connected graph, 
%\begin{align}
    %\left|\hat{\mathbf{x}}\right|=\left|\mathbf{h}^{(i)}_{\mathcal{T}}\right|
%\end{align}

%\bibliographystyle{IEEEtran}
%\bibliography{IEEEabrv,Ref_Dictionary}

\end{document}